\documentclass{aastex63}
\usepackage{amsmath}
\usepackage{gensymb}
\usepackage{multirow}

\newcommand\chiron{\textit{CHIRON}}
\newcommand\exoplanet{\textsf{exoplanet}}
\newcommand\harps{\textit{HARPS}}
\newcommand\jwst{\textit{JWST}}
\newcommand\lsun{L$_\sun$}
\newcommand\mearth{M$_\earth$}
\newcommand\msun{M$_\sun$}
\newcommand\No{\mathcal{N}}
\newcommand\planetname{TOI 540 b}
\newcommand\prot{P_{\rm rot}}
\newcommand\pymc{\textsf{PyMC3}}
\newcommand\rearth{R$_\earth$}
\newcommand\rosat{\textit{ROSAT}}
\newcommand\rsun{R$_\sun$}
\newcommand\searth{S$_\earth$}
\newcommand\starname{TOI 540}
\newcommand\teq{T_{\rm eq}}
\newcommand\tess{\textit{TESS}}
\newcommand\Un{\mathcal{U}}
\newcommand\xmm{\textit{XMM-Newton}}
\renewcommand*{\thefootnote}{\fnsymbol{footnote}}

\accepted{September 25, 2020}
\submitjournal{AJ}

\shorttitle{A Terrestrial Planet Orbiting a Nearby Rapidly Rotating Star}
\shortauthors{Ment et al.}

\graphicspath{{./}{figures/}}

\begin{document}

\title{\planetname: A Planet Smaller than Earth Orbiting a Nearby Rapidly Rotating Low-mass Star}

\correspondingauthor{Kristo Ment}
\email{kristo.ment@cfa.harvard.edu}

\author[0000-0001-5847-9147]{Kristo Ment}
\affiliation{Center for Astrophysics \textbar~Harvard \& Smithsonian, 60 Garden Street, Cambridge, MA 02138, USA}

\author{Jonathan Irwin}
\affil{Center for Astrophysics \textbar~Harvard \& Smithsonian, 60 Garden Street, Cambridge, MA 02138, USA}

\author[0000-0002-9003-484X]{David Charbonneau}
\affil{Center for Astrophysics \textbar~Harvard \& Smithsonian, 60 Garden Street, Cambridge, MA 02138, USA}

\author[0000-0001-6031-9513]{Jennifer G. Winters}
\affil{Center for Astrophysics \textbar~Harvard \& Smithsonian, 60 Garden Street, Cambridge, MA 02138, USA}

\author[0000-0001-8726-3134]{Amber Medina}
\affil{Center for Astrophysics \textbar~Harvard \& Smithsonian, 60 Garden Street, Cambridge, MA 02138, USA}

\author[0000-0001-5383-9393]{Ryan Cloutier}
\affil{Center for Astrophysics \textbar~Harvard \& Smithsonian, 60 Garden Street, Cambridge, MA 02138, USA}

\author{Mat\'{i}as R. D\'{i}az}
\affil{Departamento de Astronom\'{i}a, Universidad de Chile, Camino El Observatorio 1515, Las Condes, Santiago, Chile}

\author{James S. Jenkins}
\affil{Departamento de Astronom\'{i}a, Universidad de Chile, Camino El Observatorio 1515, Las Condes, Santiago, Chile}
\affil{Centro de Astrof\'isica y Tecnolog\'ias Afines (CATA), Casilla 36-D, Santiago, Chile}

\author{Carl Ziegler}
\affil{Dunlap Institute for Astronomy and Astrophysics, University of Toronto, 50 St. George Street, Toronto, Ontario M5S 3H4, Canada}


\author{Nicholas Law}
\affil{Department of Physics and Astronomy, The University of North Carolina at Chapel Hill, Chapel Hill, NC 27599-3255, USA}

\author[0000-0003-3654-1602]{Andrew W. Mann}
\affil{Department of Physics and Astronomy, The University of North Carolina at Chapel Hill, Chapel Hill, NC 27599-3255, USA}

\author[0000-0003-2058-6662]{George Ricker}
\affil{Department of Earth, Atmospheric and Planetary Sciences, Massachusetts Institute of Technology, Cambridge, MA 02139, USA}
\affil{Kavli Institute for Astrophysics and Space Research, Massachusetts Institute of Technology, Cambridge, MA 02139, USA}

\author{Roland Vanderspek}
\affil{Department of Earth, Atmospheric and Planetary Sciences, Massachusetts Institute of Technology, Cambridge, MA 02139, USA}
\affil{Kavli Institute for Astrophysics and Space Research, Massachusetts Institute of Technology, Cambridge, MA 02139, USA}

\author[0000-0001-9911-7388]{David W. Latham}
\affil{Center for Astrophysics \textbar~Harvard \& Smithsonian, 60 Garden Street, Cambridge, MA 02138, USA}

\author{Sara Seager}
\affil{Department of Earth, Atmospheric and Planetary Sciences, Massachusetts Institute of Technology, Cambridge, MA 02139, USA}
\affil{Kavli Institute for Astrophysics and Space Research, Massachusetts Institute of Technology, Cambridge, MA 02139, USA}
\affil{Department of Aeronautics and Astronautics, MIT, 77 Massachusetts Avenue, Cambridge, MA 02139, USA}
\affil{Department of Physics, Massachusetts Institute of Technology, Cambridge, MA 02139, USA}

\author[0000-0002-4265-047X]{Joshua N. Winn}
\affil{Department of Astrophysical Sciences, Princeton University, Princeton, NJ 08544, USA}

\author[0000-0002-4715-9460]{Jon M. Jenkins}
\affil{NASA Ames Research Center, Moffett Field, CA 94035, USA}



\author{Robert F. Goeke}
\affil{Kavli Institute for Astrophysics and Space Research, Massachusetts Institute of Technology, Cambridge, MA 02139, USA}


\author[0000-0001-8172-0453]{Alan M. Levine}
\affil{Kavli Institute for Astrophysics and Space Research, Massachusetts Institute of Technology, Cambridge, MA 02139, USA}

\author[0000-0002-0149-1302]{B\'{a}rbara Rojas-Ayala}
\affil{Instituto de Alta Investigaci\'{o}n, Universidad de Tarapac\'{a}, Casilla 7D, Arica, Chile}

\author[0000-0002-4829-7101]{Pamela~Rowden}
\affil{School of Physical Sciences, The Open University, Milton Keynes MK7 6AA, UK}

\author[0000-0002-8219-9505]{Eric B. Ting}
\affil{NASA Ames Research Center, Moffett Field, CA 94035, USA}

\author[0000-0002-6778-7552]{Joseph D. Twicken}
\affil{NASA Ames Research Center, Moffett Field, CA 94035, USA}
\affil{SETI Institute, Mountain View, CA 94043, USA}



\begin{abstract}

We present the discovery of \planetname, a hot planet slightly smaller than Earth orbiting the low-mass star 2MASS J05051443-4756154. The planet has an orbital period of $P = 1.239149$ days ($\pm$ 170 ms) and a radius of $r = 0.903 \pm 0.052$ \rearth, and is likely terrestrial based on the observed mass-radius distribution of small exoplanets at similar insolations. The star is 14.008 pc away and we estimate its mass and radius to be $M = 0.159 \pm 0.014$ \msun~and $R = 0.1895 \pm 0.0079$ \rsun, respectively. \replaced{It}{The star} is distinctive in its very short rotational period of $\prot = 17.4264 +/- 0.0094$ hours and correspondingly small Rossby number of 0.007 as well as its high X-ray-to-bolometric luminosity ratio of $L_X / L_{\rm bol} = 0.0028$ based on \replaced{an}{a serendipitous} \xmm~detection during a slew operation. This is consistent with the X-ray emission being observed at a maximum value of $L_X / L_{\rm bol} \simeq 10^{-3}$ as predicted for the most rapidly rotating M dwarfs. \planetname~may be an alluring target to study atmospheric erosion due to the strong stellar X-ray emission. It is also among the most accessible targets for transmission and emission spectroscopy and eclipse photometry with \jwst, and may permit Doppler tomography with high-resolution spectroscopy during transit. This discovery is based on \deleted{highly} precise photometric data from \tess~and ground-based follow-up observations by the MEarth team.

\end{abstract}


\section{Introduction} \label{sec:intro}

Obtaining a representative sample of rotation periods in planet-hosting M dwarfs is important due to the established link between rotation rate and stellar activity. Stars exhibit continual angular momentum loss due to magnetic braking, which subsequently decreases the strength of the stellar magnetic dynamo, leading to a decrease in activity \citep{Skumanich1972}. Consequently, rapidly rotating stars tend to have higher levels of coronal X-ray emission \citep{Wright2018}. Elevated X-ray and UV emission are known tracers of increased magnetic activity \citep{Gronoff2020} and will have profound consequences on the atmospheric mass loss and potential habitability of any terrestrial planets in the system \citep[e.g.][]{GarciaSage2017}.

The vast majority of recent discoveries of terrestrial planets orbiting M dwarfs are in systems with relatively inactive host stars, characterized by stellar rotation periods longer than 100 days. Rapidly rotating stars present observational challenges for planet detection. Rotational broadening of spectral lines significantly degrades radial velocity (RV) precision, and irregularities of emitted flux from starspots and faculae that are rapidly shifting in and out of view across the stellar surface can imprint complicated modulations in the star's observed flux, hampering transit detection. In addition, rapid rotation in M dwarfs also correlates with more frequent flare emission \citep[and references therein]{Davenport2016}, further contaminating the light curve. In particular, the current sample of terrestrial planet hosts within 15 pc of the Sun includes no red dwarfs with rotation periods less than a day, and only one ultracool dwarf with a rotation period below 10 days: TRAPPIST-1, which has an estimated $\prot = 1.40 \pm 0.05$ days \citep{Gillon2016} or $\prot = 3.30 \pm 0.14$ days \citep{Luger2017}.

Transiting planets with radii around or less than 1\rearth~are of special interest since they are likely to be terrestrial even in the absence of mass constraints from precise RV or transit-timing variation (TTV) data. \citet{Weiss2014} and \citet{Rogers2015} identified two distinct regimes in the mass-radius relation for small exoplanets: planets below a threshold radius of 1.5\rearth~tend to have bulk densities consistent with a rocky composition whereas those above that threshold possess a large fraction of volatiles by volume. These results were based on numerous RV studies that had obtained masses for small transiting planets around G and K dwarfs such as Kepler-10 \citep{Batalha2011}, Kepler-78 \citep{Howard2013,Pepe2013}, and Kepler-93 \citep{Dressing2015a}. Recent RV measurements have established rocky bulk composition for many small planets orbiting M dwarfs as well, including L 98-59 \citep{Cloutier2019}, LHS 1140 \citep{Ment2019}, LTT 3780 \citep{Cloutier2020a}, and TRAPPIST-1 \citep{Grimm2018}. Consistent with the interpretation of two distinct planet populations, \citet{Fulton2017} noticed a deficit of planets with radii 1.5-2\rearth~(a "radius valley"). The radius valley likely arises as a result of photoevaporation \citep{Lopez2013,Owen2013,Owen2017,Lopez2018}, core-powered mass loss \citep{Ginzburg2018}, or formation in a gas-poor environment \citep{Lee2014,Lopez2018}. Its existence was demonstrated specifically for K and M dwarfs by \citet{Cloutier2020} using K2 photometry.

Upcoming exo-atmospheric studies with state-of-the-art instruments such as the \textit{James Webb Space Telescope} (\jwst) and ground-based 30-meter telescopes will primarily focus on the planets orbiting M dwarfs closest to the Solar System. This is due to photon statistics as well as the relative size of the planet compared to the star that make both transmission and emission spectroscopy tractable \citep{Morley2017}. There are currently (as of April 2020) only ten stars within 15 pc of the Sun that are known to host transiting planets: the M dwarfs LTT 1445 A \citep{Winters2019}, GJ 357 \citep{Luque2019}, GJ 436 \citep{Butler2004,Gillon2007}, L 98-59 \citep{Kostov2019}, TRAPPIST-1 \citep{Gillon2016,Gillon2017}, GJ 1132 \citep{BertaThompson2015,Bonfils2018}, GJ 1214 \citep{Charbonneau2009}, LHS 3844 \citep{Vanderspek2019}, LHS 1140 \citep{Dittmann2017,Ment2019}, and the K dwarfs 55 Cnc A \citep{McArthur2004,Winn2011} and HD 219134 \citep{Motalebi2015,Vogt2015,Gillon2017a}; only two of those systems (L 98-59 and TRAPPIST-1) are known to host planets smaller than 1\rearth. Four of the planetary systems described above (LTT 1445, GJ 357, L 98-59, and LHS 3844) were discovered with the \textit{Transiting Exoplanet Survey Satellite} \citep[\tess;][]{Ricker2015}.

This article presents the discovery of \tess~Object of Interest (TOI) 540 b, a 0.9\rearth~planet on a 1.24-day orbit discovered by \tess. With a rotation period of 17.4 hours, \starname~is rotating more rapidly than TRAPPIST-1, the only other rapidly rotating planet host within 15 pc. We complement \tess~photometry with ground-based follow-up photometry from MEarth, confirming the planetary nature of the candidate. In addition, we present RV measurements from \chiron~and \harps, speckle imaging from SOAR, and X-ray data from \xmm~and \rosat.

\section{Properties of the host star}\label{sec:star}

\starname, otherwise known as 2MASS J05051443-4756154 \citep{Cutri2003,Skrutskie2006} and UCAC4 211-005570 \citep{Finch2014}, is a nearby main-sequence M dwarf. Based on the parallax measurement of $\pi = 71.3886 \pm 0.0448~\rm{mas}$ reported in Gaia DR2 \citep{Gaia2018,Lindegren2018}, we calculate a distance of $d = 14.0078 \pm 0.0088$ pc from Earth. \starname~has a low proper motion for a nearby star ($\mu_\alpha = -66.09 \pm 0.08$ mas yr$^{-1}$, $\mu_\delta = 25.08 \pm 0.09$ mas yr$^{-1}$) which likely prevented it from being widely identified as a nearby star before Gaia. The proper motion is also too low to enable ruling out the presence of background stars from archival images (in Section \ref{sec:speckle}, we present high-angular-resolution images that rule out the presence of other bright stars in the immediate vicinity). To obtain an estimate for the mass of the star, we use the mass-luminosity relationship for main-sequence M dwarfs in \citet{Benedict2016} and the $K$-band apparent magnitude of 8.900 $\pm$ 0.021 from 2MASS. This yields a stellar mass of $M = 0.159 \pm 0.014$ \msun. We determine the stellar radius by using two different mass-radius relations: one determined from optical interferometry of single stars in \citet{Boyajian2012}, and another from eclipsing binary measurements in \citet{Bayless2006}. The former yields $R = 0.195 \pm 0.011$ \rsun, and the latter produces $R = 0.182 \pm 0.013$ \rsun. We then calculate a weighted average of the two estimates, obtaining $R = 0.190 \pm 0.008$ \rsun. We note that this is also consistent with the radius-luminosity relation in \citet{Mann2015}, which predicts $R = 0.197 \pm 0.007$ \rsun. Informed by these relationships, we refine our estimate of the stellar radius using transit geometry in Section \ref{sec:modeling} and report the final value in Table \ref{tbl:results}.

In order to determine the color indices of \starname, we adopt the $J$-, $H$-, and $K_S$-band magnitudes from 2MASS \citep{Skrutskie2006}, and we obtained $V_{J}R_{KC}I_{KC}$-photometry from RECONS\footnote{REsearch Consortium On Nearby Stars (\url{www.recons.org})}. The $VRI$-photometry was collected on the night of 20 Aug 2019, with exposure times of 300, 180, and 75 seconds, respectively. The RECONS fluxes were extracted using a 4\arcsec~radius aperture to minimize contamination from a nearby background star, described in Section \ref{sec:bgstar}. The VRI magnitudes for \starname~are reported in Table \ref{tbl:results}.

The luminosity of \starname~can be estimated from bolometric corrections ($BC$). In particular, interpolating between the $BC_V$ values as a function of $V-K_S$ in Table 5 of \citet{Pecaut2013} yields a bolometric correction of $BC_V = -2.938$, corresponding to a luminosity of $L = 0.003745$ \lsun. Alternatively, using the derived third-order polynomial fit between $BC_J$ and $V-J$ in \citet[and its erratum]{Mann2015}, we obtain $BC_J = 1.951$, which gives us a luminosity of $L = 0.003254$ \lsun. Finally, we use the relationship between $BC_K$ and $I-K$ in \citet{Leggett2001} to produce $BC_K = 2.764$ and $L = 0.003383$ \lsun. We take as our final value the mean and the standard deviation of the three luminosity estimates, $L = 0.00346 \pm 0.00021$ \lsun. This allows us to use the Stefan-Boltzmann law to determine the effective stellar temperature via $T_{\rm eff} = T_{\rm eff,\sun} (L/L_\sun)^{1/4} (R/R_\sun)^{-1/2}$, yielding $T_{\rm eff} = 3216 \pm 83$ K. We adopted the solar values of $M_{\rm bol,\sun} = 4.7554$ mag and $T_{\rm eff,\sun} = 5772$ K cited in \citet{Mamajek2012}.

Due to the slow evolution of M dwarfs in their rapid rotation stage, we are unable to place tight constraints on the age of the system. However, we note that \starname~is not overluminous (based on the observed color and luminosity), and therefore it is likely to be on the main sequence, implying an age greater than 100 Myr \citep{Baraffe2002}. Based on the galactic space velocity $W$ and established age-velocity relations, and the observed rapid stellar rotation, we can place an upper limit of 2 Gyr on the system \citep{Newton2016}. We also estimate the average equivalent width of H$\alpha$ emission from the four \chiron~spectra described in Section \ref{sec:spectra} to be $2.8 \pm 0.1$ \AA, consistent with \starname~not being a pre-main sequence star. Finally, we use BANYAN $\Sigma$ \citep{Gagne2018} to investigate \starname's potential membership of young stellar associations based on its location in XYZUVW space (calculated from the coordinates, proper motion, and RV values), and rule out \deleted{all} 27 well-characterized young associations within 150 pc with 99.9\% confidence.

\starname~is a highly magnetically active star \replaced{exemplified}{as indicated} by the photometric modulation due to stellar spots and the numerous flares present in each sector of TESS data. Measuring how often \starname~flares is essential to understanding the environment in which its planet resides. The flare frequency distribution describes the rate of flares as function of energy and follows the probability distribution $N(E)dE = \Omega E^{-\alpha} dE$ \citep[equation 2]{Lacy1976} where $\alpha$ is the slope of the power law and $\Omega$ is a normalization constant. Using the methods outlined in Medina et al. (submitted\footnote{Medina, A. A., Winters, J. G., Irwin, J. M., \& Charbonneau, D., submitted}), we find $\alpha = 1.97$ and measure a rate of 0.12 flares per day above an energy of $E = 3.16 \times 10^{31}$ ergs in the TESS bandpass. Medina et al. find that \starname~has a flare rate that is consistent with other stars of a similar mass and rotation period.

\subsection{A neighboring star 6\arcsec~away}\label{sec:bgstar}

Our efforts are complicated by a neighboring background star at a distance of 6\arcsec~from \starname. We first noticed this background source in MEarth follow-up photometry (described in Section \ref{sec:mearth}) and subsequently identified it in Gaia DR2. It has a Gaia ID of 4785886975670558336 and is 2.26 magnitudes fainter than the main target in the $G_{BP}$ passband. While both the neighboring star and \starname~occupy the same \tess~pixel, we are able to resolve the two stars using MEarth photometry and confirm that the planet does indeed transit \starname. In particular, we extract the light curve of \starname~using variable aperture sizes and find that the transit persists with a similar depth down to an aperture of 2.5\arcsec, a size small enough to exclude the neighboring star.

\section{Observations}

\subsection{\tess~photometry}\label{sec:tess}

\tess~collected photometry of \starname~in observation sectors 4, 5, and 6, with the observations spanning a nearly three-month period from 19 October 2018 to 6 January 2019. The star was included in the \tess~Input Catalog (TIC) with a TIC ID of 200322593 as well as the \tess~Candidate Target List \citep[CTL;][]{Stassun2018}, and \tess~Guest Investigator programs G011180 (PI: Courtney Dressing) and G011231 (PI: Jennifer Winters). We utilize the two-minute cadence Presearch Data Conditioning \citep[PDCSAP;][]{Smith2012,Stumpe2012,Stumpe2014} light curve reduced with the NASA Ames Science Processing Operations Center (SPOC) pipeline \citep{Jenkins2016}. The PDCSAP fluxes have been corrected for instrumental systematic effects as well as crowding: unresolved light from other nearby stars listed in the TIC v7. While the background source from Section \ref{sec:bgstar} is listed in version 8 of TIC, it did not appear in version 7 that was used to reduce the \tess~photometry presented in this work. Based on TIC v8, the background star is 4.46 magnitudes fainter than \starname~in the TESS bandpass. This leads to an error of 1.6\% in the measured transit depth, or 0.8\% in the planetary radius. The latter is much smaller than the final uncertainty of 5.8\% derived in this work (Table \ref{sec:results}). Therefore, we have not corrected the \tess~light curve presented in Section \ref{sec:tess} for the additional flux dilution due to this neighboring source.

A planetary candidate with an orbital period of 1.239 days was initially detected by SPOC in sector 4 Data Validation Reports \citep[DVR;][]{Twicken2018,Li2019} with a signal-to-noise ratio of 8.9, based on 16 transits. \tess~ultimately observed 50 transits of \planetname~over the three sectors, with an average transit depth of 2168 $\pm$ 172 ppm, yielding a signal-to-noise ratio of 15.6. However, the correct spectral type and stellar radius was undefined in the TIC (v7) and therefore assumed to be 1\rsun~during the preparation of the DVR, leading to a substantially overestimated planetary radius of 5\rearth. 

\begin{figure}
    \includegraphics[width=\textwidth]{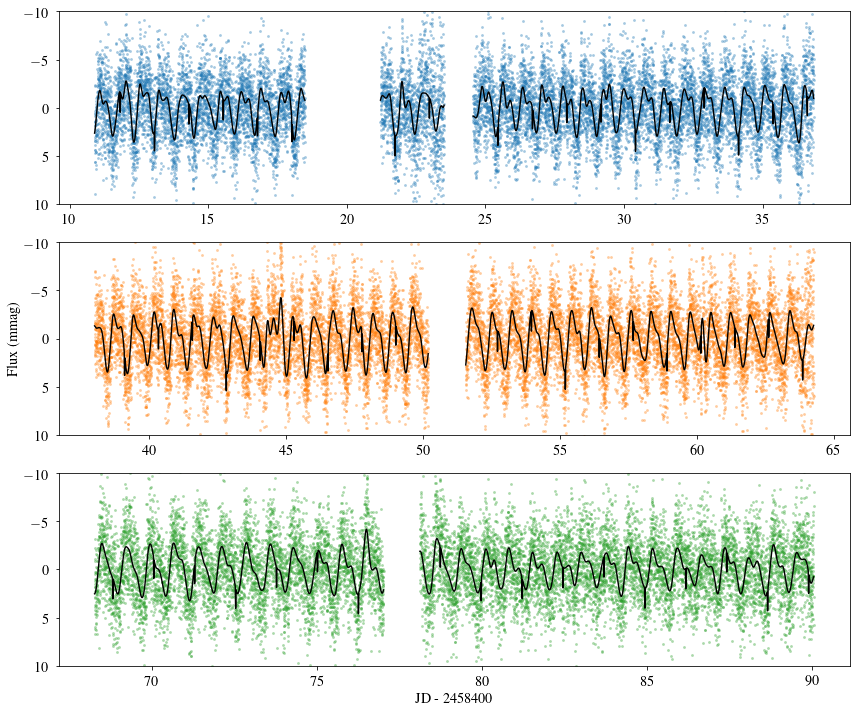}
    \caption{TESS photometry of \starname~from Sectors 4 (blue), 5 (orange), and 6 (green). The combined transit and stellar rotation model from Section \ref{sec:modeling} is overplotted as a solid line.}
    \label{fig:tessgp}
\end{figure}

We make use of the full PDCSAP light curve from which we remove bright outliers (more than 0.02 mag brighter than the mean flux) that constitute 0.099\% of the total \tess~data and may be caused by contamination from flares. The data clipping is done to ensure consistency between the handling of \tess~and MEarth data sets (see Section \ref{sec:mearth}). The final \tess~light curve consists of 48,445 individual data points, and can be seen in Figure \ref{fig:tessgp}. We note that per the \tess~Data Release Notes of Sector 4\footnote{\url{https://archive.stsci.edu/missions/tess/doc/tess\_drn/tess\_sector\_04\_drn05\_v04.pdf}}, an interruption in communications between the instrument and spacecraft resulted in an instrument turn-off for 2.7 days, during which no data were collected.

\subsection{MEarth follow-up photometry}\label{sec:mearth}

\begin{figure}
    \includegraphics[width=\textwidth]{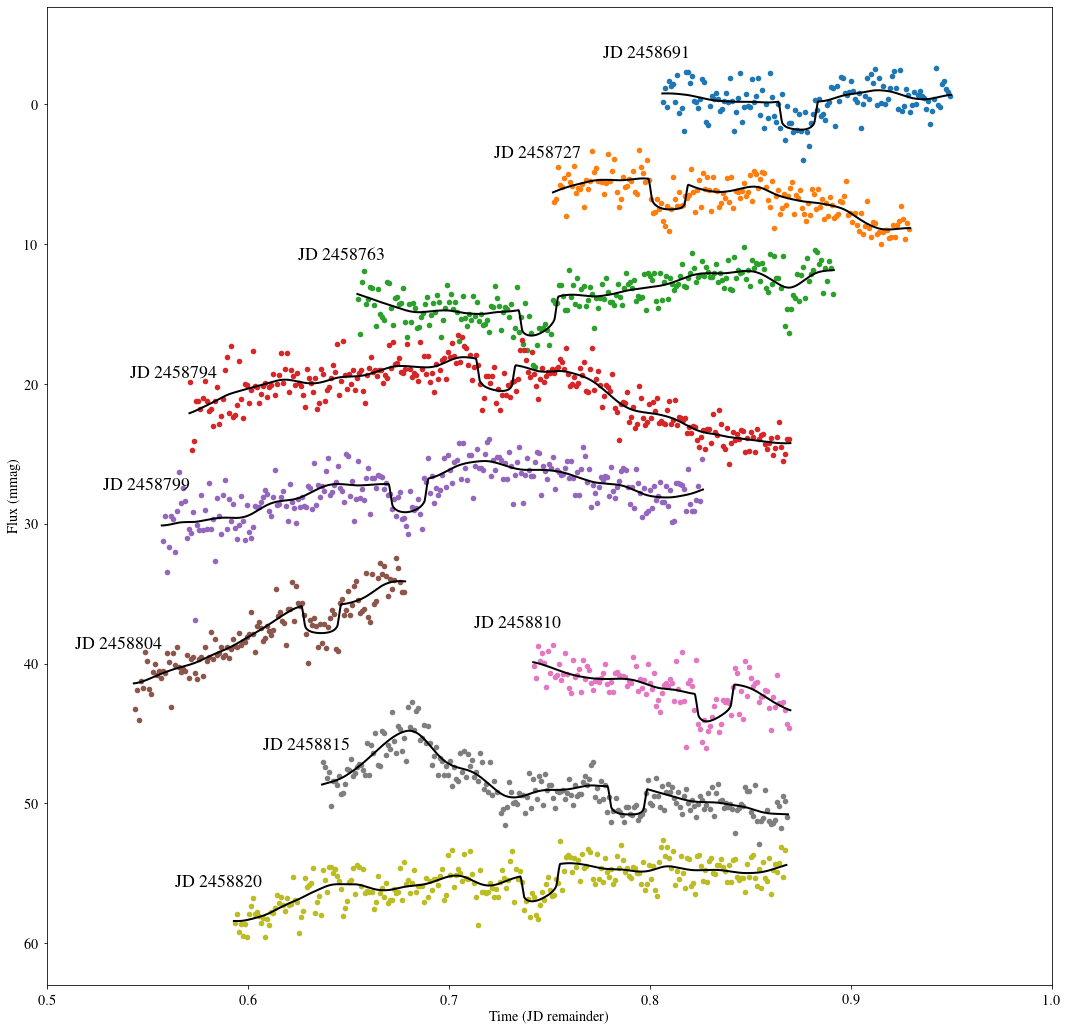}
    \caption{Follow-up photometry of \starname~from all 8 MEarth telescopes by night. The data has been combined and binned to a cadence of 1.44 minutes. The solid line represents the combined transit and baseline model from Section \ref{sec:modeling}.}
    \label{fig:mearth}
\end{figure}

Using the eight 40cm aperture telescopes of the MEarth-South telescope array at the Cerro Tololo International Observatory (CTIO) in Chile \citep{Nutzman2008,Irwin2015}, we conducted follow-up observations of \starname~to confirm the transits of the terrestrial planet. The MEarth-South telescopes employ a custom bandpass centered at the red end of the optical spectrum (similar to \tess). We observed 10 transits of \planetname~between 27 July 2019 and 8 December 2019, using an exposure time of 40 seconds per measurement. However, we discarded the data from the transit on Dec 8th due to a stellar flare shortly following the transit egress, which would have led to needless challenges in modeling the out-of-transit flux baseline. Therefore, we proceeded by including the data from the first 9 transits only. Aperture photometry was carried out in all images using a fixed aperture radius of 6 pixels, or 5.1\arcsec. After excluding outliers 0.02 mag brighter than the mean flux (0.068\% of the data set), the MEarth light curve contains a combined 17,879 data points from all eight telescopes. The individual light curves from each visit are shown in Figure \ref{fig:mearth}.

\subsection{\chiron~and \harps~spectroscopy}\label{sec:spectra}

\begin{figure}
    \includegraphics[width=\textwidth]{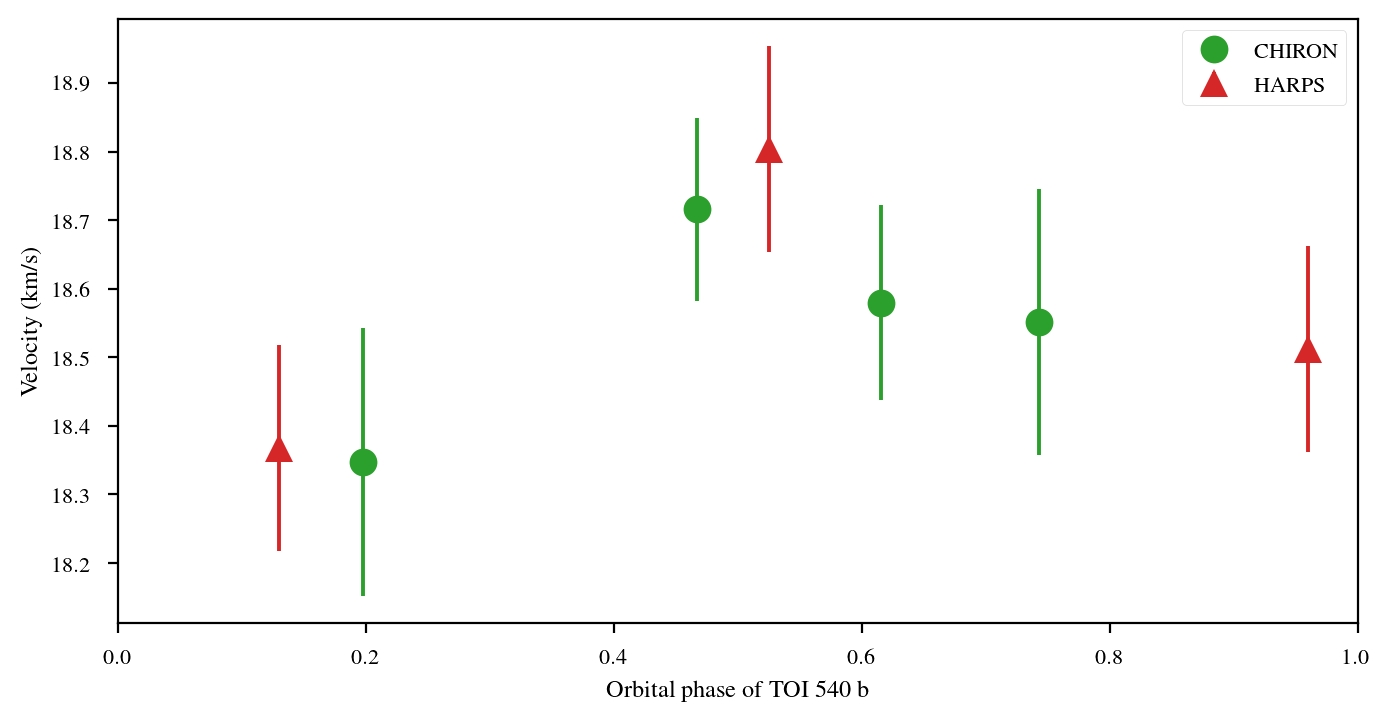}
    \caption{RV measurements of \starname~from \chiron~and \harps~as a function of orbital phase, with phase of zero corresponding to the time of transit. There is no evidence for a significant trend caused by a massive companion. The expected RV semi-amplitude for \planetname~is $K = 1.4 \pm 0.3$ m/s (assuming a circular orbit), well below the precision of the currently available data.}
    \label{fig:rv}
\end{figure}

We gathered 4 reconnaissance spectra of \starname~with the \chiron~spectrograph \citep{Tokovinin2013} mounted on the CTIO/SMARTS 1.5-meter telescope at CTIO. The spectra were accumulated as part of a nearly volume-complete spectroscopic survey of nearby mid-to-late M dwarfs. The methods by which we determined the radial velocities and rotation broadening, below, are described in \citet{Winters2020}. The observations were carried out between Sep 2018 and Nov 2019 in 3x20-minute exposures per observation, employing the image slicer mode for a resolution of $R = 80,000$. We obtained multi-order RVs from 6 spectral orders (see Table \ref{tbl:rv}) as well as an estimated projected rotation velocity of $v \sin i = 13.98$ km s$^{-1}$. In addition, we collected 3 spectra using the \harps~spectrograph \citep{Mayor2003} mounted on the ESO 3.6-meter telescope at La Silla. The \harps~spectrograph has a measured spectral resolution of $R = 120,000$. The observations were carried out in April 2019 (ESO \harps~Program 0103.C-0442, PI: D\'{i}az) with a 30-minute exposure time. We estimated the RVs by combining 21 spectral orders and calculated a $v \sin i$ of 12.93 km/s. The RVs are displayed in Figure \ref{fig:rv}.

\begin{deluxetable}{cccc}
    \tabletypesize{\footnotesize}
	\tablewidth{0pt}
	\tablecaption{RV data for \starname\label{tbl:rv}}
	\tablehead{
		BJD (TDB) & RV & Uncertainty & Instrument\\
		 & (km/s) & (km/s) &
	}
	\startdata
	2458385.8447 & 18.5791 & 0.142 & \chiron\\
	2458576.5615 & 18.9040 & 0.1886 & \harps\\
	2458578.5499 & 18.4817 & 0.1375 & \harps\\
	2458579.5788 & 18.6201 & 0.1221 & \harps\\
	2458607.4689 & 18.7156 & 0.133 & \chiron\\
	2458801.6815 & 18.3476 & 0.195 & \chiron\\
	2458803.5955 & 18.5508 & 0.194 & \chiron\\
	\enddata
\end{deluxetable}

We adopt the mean of the two rotation velocity estimates as our final value, obtaining $v \sin i = 13.5 \pm 1.5$ km s$^{-1}$. This yields a rotation period estimate of $\frac{\prot'}{\sin i} = \frac{2\pi R}{v \sin i} = 0.71 \pm 0.08$ days, consistent with the photometrically-determined rotation period of 0.72610 days from Section \ref{sec:modeling} for $\sin i \approx 1$, suggesting that the sky-projected stellar obliquity is likely close to 90\degree. \deleted{The uncertainty in $v \sin i$ was approximated from the standard deviation of the $v \sin i$ values from 6 individual spectral orders in the four \chiron~spectra.} The $v \sin i$ values \added{for both \chiron~and \harps~} were generated by applying appropriate rotational broadening to an observed M dwarf spectrum.

We note that our RV measurements have uncertainties greater than 100 m s$^{-1}$, estimated from theoretical uncertainties for a rotating star \citep[e.g.][]{Bouchy2001} with inflation to account for RV scatter in between the spectral orders. The large uncertainties are driven by the low signal-to-noise ratios of the spectra (typically 5-10) due to the star being substantially redder than what is optimal for the spectrographs and settings that were used to collect these spectra. The \harps~spectra were gathered in the Simultaneous Reference mode using a Thorium-Argon lamp which significantly degrades the spectrum in the red orders. Rotational broadening of spectral lines also contributes to the degradation of RV precision. The reported errors are on a par with the RMS of the RVs (163 m s$^{-1}$).
\explain{Edited for clarity.}

Employing the estimated mass of the planet from Section \ref{sec:results}, the RV semi-amplitude for \planetname~corresponding to a circular orbit would be $K = 1.4 \pm 0.3$ m/s. Therefore, the rapid rotation of the star makes direct mass measurements with RVs currently unfeasible. However, the RVs help rule out a close-in massive companion: the standard deviation of all 7 measurements is 167 m/s, close to the individual uncertainties, and there is no evidence of a trend. In particular, we are able to rule out a companion with a mass of $m > 0.73 M_{\rm Jup}$ at the orbital period of the transiting planet with 99.7\% (3-sigma) confidence. This was estimated by fitting an RV model with the period, epoch, and eccentricity fixed to the values in Table \ref{tbl:results}, and calculating the appropriate semi-amplitude for which the CDF of the $\chi^2$ distribution has a $p$-value below 0.0027.

\subsection{SOAR speckle imaging}\label{sec:speckle}

\begin{figure}
    \includegraphics[width=\textwidth]{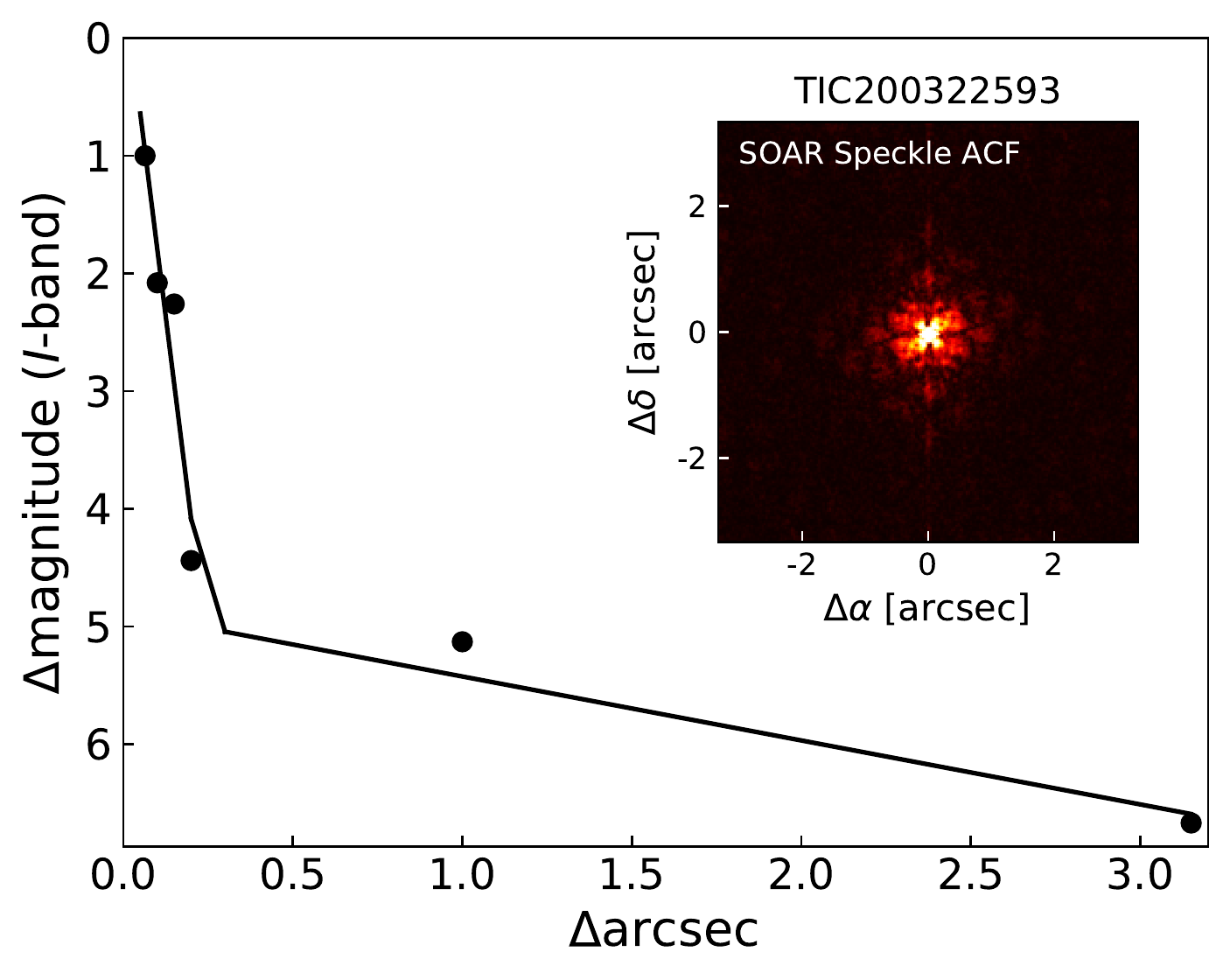}
    \caption{The 5$\sigma$ detection sensitivity (solid line) and the auto-correlation function (ACF, inset) of \starname~from SOAR speckle imaging showing no evidence of nearby light sources.}
    \label{fig:speckle}
\end{figure}

Nearby stars that fall within the same 21\arcsec~\tess~pixel as the target can cause photometric contamination or be the source of an astrophysical false positive. We searched for nearby sources to \starname~with SOAR speckle imaging \citep{Tokovinin2018} on 14 July 2019 UT, observing in the visible $I$-bandpass. Details of the observation are available in \citet{Ziegler2020}. We detected no nearby sources within 3\arcsec~of \starname, corresponding to a projected distance of 42 AU. The 5-sigma detection sensitivity and the speckle auto-correlation function (contrast curve) from the SOAR observation are plotted in Figure \ref{fig:speckle}.

\subsection{X-ray detections by \xmm~and \rosat}

A testament to its significant X-ray brightness, \starname~was detected by \xmm~during a slew operation on 27 July 2004. The detection appears in the XMM Slew 2 catalog with a source ID of XMMSL2 J050514.2-475618 (Slew Obs. ID 9084800002). The flux of the target is listed as $(1.43 \pm 0.50) \times 10^{-12}$ ergs s$^{-1}$ cm$^{-2}$ in the soft 0.2-2 keV bandpass and $(3.91 \pm 1.35) \times 10^{-12}$ ergs s$^{-1}$ cm$^{-2}$ in the total 0.2-12 keV bandpass. No detection is listed separately in the hard 2-12 keV bandpass. For consistency with \citet{Wright2018}, we convert the soft bandpass flux into the \rosat~bandpass of 0.1-2.4 keV. We use the APEC\footnote{Astrophysical Plasma Emission Code, \url{http://www.atomdb.org/}} model at solar abundance and a plasma temperature of $kT = 0.54$ keV. The conversion is done using PIMMS\footnote{The Portable Interactive Multi-Mission Simulator, \url{https://heasarc.gsfc.nasa.gov/docs/software/tools/pimms.html}} and yields an X-ray flux of $(1.60 \pm 0.56) \times 10^{-12}$ ergs s$^{-1}$ cm$^{-2}$ in the \rosat~bandpass. This corresponds to an X-ray-to-bolometric luminosity ratio of $L_X / L_{\rm bol} = 0.0028$, consistent with $L_X / L_{\rm bol} \simeq 10^{-3}$ measured by \citet{Wright2018} for rapidly rotating stars. The implications of this are further discussed in Section \ref{sec:results}.

\starname~also appears in the second ROSAT all-sky survey source catalog \citep{Boller2016} with a source name of J050514.2-475625. It was observed by \rosat~in August 1990 with a count rate of $0.0413 \pm 0.0176$ counts s$^{-1}$ in the 0.1-2.4 keV energy band. Using the aforementioned APEC model leads to a somewhat lower X-ray flux estimate of $3 \times 10^{-13}$ ergs s$^{-1}$ cm$^{-2}$; however, the error on the count rate is substantial.

\section{Modeling of \tess~and MEarth light curves}\label{sec:modeling}

We analyze the MEarth and TESS light curve data simultaneously with the Python package \exoplanet~\citep{exoplanet}, which is a framework built on the Hamiltonian Monte Carlo methods implemented in \pymc~\citep{exoplanet:pymc3} via \textsf{Theano}~\citep{exoplanet:theano} for computationally efficient sampling. Importantly, \exoplanet~extends the basic support for Gaussian Process (GP) modeling in \pymc~by implementing Scalable GPs through \textsf{celerite}~\citep{exoplanet:foremanmackey18}, which makes the otherwise notoriously slow GP modeling much more tractable. Our model has multiple components that are optimized simultaneously, described in the following sections.

We model the rotational modulation in the TESS photometry with a GP employing \exoplanet's Rotation kernel, which has a covariance function that is a sum of two stochastically-driven harmonic oscillators (SHO):
\begin{equation}\label{eq:ktess}
    k_{\rm TESS}(\tau; Q_0, \Delta Q, \prot, \mathcal{A}, \lambda) = k_{\rm SHO}(\tau; Q_1, \omega_1, S_1) + k_{\rm SHO}(\tau; Q_2, \omega_2, S_2)
\end{equation}
\begin{align}
    Q_1 = \frac{1}{2} + Q_0 + \Delta Q && Q_2 = \frac{1}{2} + Q_0 \nonumber\\
    \omega_1 = \frac{4\pi Q_1}{\prot\sqrt{4Q_1^2 - 1}} && \omega_2 = \frac{8\pi Q_2}{\prot\sqrt{4Q_2^2 - 1}} \nonumber\\
    S_1 = \frac{\mathcal{A}}{\omega_1 Q_1} && S_2 = \frac{\lambda\mathcal{A}}{\omega_2 Q_2} \nonumber
\end{align}
where the covariance function of a single SHO is given by:
\begin{equation}\label{eq:ksho}
k_{\rm SHO} (\tau; Q, \omega, S) = S \omega Q e^{-\frac{\omega\tau}{2Q}} 
    \begin{cases}
        \cosh(\eta\omega\tau) + \frac{1}{2\eta Q} \sinh(\eta\omega\tau) & \rm if~0 < Q < 1/2 \\
        2 (1 + \omega\tau) & \rm if~Q = 1/2 \\
        \cos(\eta\omega\tau) + \frac{1}{2\eta Q} \sin(\eta\omega\tau) & \rm if~Q > 1/2
    \end{cases}
\end{equation}
with $\eta \equiv \sqrt{\lvert 1 - \frac{1}{4 Q^2} \rvert}$. Thus, the rotation kernel has five parameters: the quality factors $Q_0$ and $\Delta Q$, the rotation period $\prot$, the primary amplitude $\mathcal{A}$, and the amplitude ratio $\lambda$. We note that the two SHO components correspond to the first and second harmonics of the oscillation, with the latter having twice the frequency (or equivalently, half the period) of the former. This type of kernel has been shown to successfully model a range of complicated rotationally modulated signals \citep{Haywood2014,Soto2018,Winters2019,Cloutier2020b}. The parameters of the kernel are constrained with normal and uniform prior distributions, documented in Table \ref{tbl:params}. We obtain an initial estimate of 17.4 hours for $\prot$ by fitting a sum of two sinusoids to the \tess~light curve, and we subsequently constrain $\prot$ with a normal prior distribution centered at that value. The prior for $\mathcal{A}$ is centered at the value corresponding to the variance of the observed TESS light curve. We note that since $Q_0$ will always be positive with this setup, the quality parameters $Q_1$ and $Q_2$ of both SHOs are guaranteed to remain above $1/2$ in Equation \ref{eq:ksho}.

The MEarth follow-up photometry is modeled with a separate covariance kernel. This is due to the additional contribution from precipitable water vapor that induces strong non-linear trends into the individual light curves of each night. The form of the kernel is as follows:
\begin{equation}\label{eq:kmearth}
    k_{\rm MEarth}(\tau; \sigma, \rho) = \frac{\sigma^2}{2} \left[ \left(1 + \frac{1}{\epsilon}\right) e^{-(1-\epsilon)\sqrt{3}\tau/\rho} + \left(1 - \frac{1}{\epsilon}\right) e^{-(1+\epsilon)\sqrt{3}\tau/\rho} \right]
\end{equation}
which, for small values of $\epsilon$, approximates the well-known Mat\'ern-3/2 kernel:
\begin{equation}\label{eq:kmatern32}
    \lim_{\epsilon \to 0} k_{\rm MEarth}(\tau; \sigma, \rho) = \sigma^2 \left(1 + \frac{\sqrt{3}\tau}{\rho}\right) e^{-\sqrt{3}\tau/\rho}
\end{equation}
Here, $\epsilon$ is fixed to a standard value of 0.01. The exact form of the Mat\'ern-3/2 kernel (Eq. \ref{eq:kmatern32}) cannot be implemented within the framework of \textsf{celerite}, and we therefore need to use the approximate form of Eq. \ref{eq:kmearth}. The timescale parameter $\rho/\sqrt{3}$ is constrained with a Gaussian prior centered at 30 minutes, which we expect to be the typical minimum timescale for variations in precipitable water vapor in the atmosphere. The prior for $\sigma$ is centered at the value corresponding to the variance of the observed MEarth light curve. We note that due to the estimated stellar rotation period of 17.4 hours, the kernel in Eq. \ref{eq:kmearth} is also able to absorb the rotational modulation with the exponential decay timescale being much shorter than the rotation period, whereas a quasi-periodic kernel (such as Eq. \ref{eq:ktess}) would be misled by the non-periodic changes in water vapor content. We did experiment with a kernel that combined both a quasi-periodic and a Mat\'ern-3/2 term, but we ultimately found it impossible to decouple the effects of stellar rotation and water vapor variations in a statistically significant way. Thus, the Mat\'ern-3/2 kernel is our preferred model to account for both periodic as well as non-periodic modulations in the high-cadence but short-baseline MEarth follow-up data.

Transits of \planetname~are modeled using the \textsf{starry} module \citep{exoplanet:luger18} included in \exoplanet. A single transit model is fitted simultaneously to TESS and MEarth photometry. While the model presented here does not account for the non-zero exposure times, we did test a model light curve that was oversampled and integrated over the different exposures (also with \exoplanet) and found the differences to be negligible. The free parameters in the model are the transit midpoint $t_0$, the orbital period $P$, the planet-to-star radius ratio $r/R$, and the impact parameter $b$. The first two ($t_0$ and $P$) are constrained relatively tightly with Gaussian priors since they can be pre-determined with good precision from the TESS data alone. The two dimensionless parameters ($r/R$ and $b$) have uniform prior distributions. We also include quadratic limb darkening, with the appropriate coefficients adopted from Table 5 of \citet{Claret2018} for the spherical PHOENIX-COND limb darkening model \citep{Husser2013}. In particular, we use $a = 0.1553$ and $b = 0.4742$, corresponding to a local gravity of $\log g = 5.0$, an effective temperature of 3200 K, and the TESS bandpass. We separately estimated the limb darkening coefficients for the MEarth optical filter following the process outlined in Section 6.1 of \citet{Irwin2018}. However, due to the similarity between the MEarth and TESS bandpasses, we found no meaningful difference between using either set of limb darkening coefficients: the resulting discrepancy in the modeled light curve was orders of magnitude smaller than the measurement uncertainty. Therefore, we decided to simplify the modeling by adopting the TESS limb darkening coefficients given above for both data sets.

In addition, we allow for a baseline flux as well as additional white noise (added in quadrature to each individual flux uncertainty) in the TESS and MEarth data as additional model parameters. They are loosely constrained with Gaussian prior distributions. The priors for excess white noise are centered at the values corresponding to the smallest individual flux uncertainties in the respective data sets. A comprehensive list of all model parameters is given in Table \ref{tbl:params}.

\begin{deluxetable}{ccccc}
    \tabletypesize{\footnotesize}
	\tablewidth{0pt}
	\tablecaption{Model parameters for light curve fitting\label{tbl:params}}
	\tablehead{
		Parameter & Explanation & Prior & Value & Units
	}
	\startdata
	$\mu_{\rm TESS}$ & TESS flux baseline & $\No(0, 10)$ & -0.025 $\pm$ 0.025 & mmag\\
	$\ln \sigma_{\rm TESS}^2$ & TESS excess white noise & $\No(\ln 6.76, 5)$ & -8.44 $\pm$ 2.04 & mmag$^2$\\
	$\mu_{\rm MEarth}$ & MEarth flux baseline & $\No(0, 10)$ & -0.49 $\pm$ 0.40 & mmag\\
	$\ln \sigma_{\rm MEarth}^2$ & MEarth excess white noise & $\No(\ln 3.44, 5)$ & 1.154 $\pm$ 0.028 & mmag$^2$\\
	$M$ & Stellar mass & $\No(0.159, 0.014)$ & Table \ref{tbl:results} & \msun\\
	$R$ & Stellar radius & $\No(0.190, 0.008)$ & Table \ref{tbl:results} & \rsun\\
	$\ln Q_0$ & Quality parameter & $\No(1, 10)$ & 1.07 $\pm$ 0.19 & -\\
	$\ln \Delta Q$ & Quality parameter & $\No(2, 10)$ & 7.88 $\pm$ 0.80 & -\\
	$\ln \prot$ & Stellar rotation period & $\No(\ln 0.7266, 0.01)$ & Table \ref{tbl:results} & days\\
	$\ln \mathcal{A}$ & Covariance amplitude & $\No(\ln 10.62, 5)$ & 0.54 $\pm$ 0.63 & mmag$^2$\\
	$\lambda$ & Covariance amp. ratio & $\Un(0, 1)$ & 0.54 $\pm$ 0.25 & -\\
	$\ln \sigma^2$ & Covariance amplitude & $\No(\ln 16.84, 5)$ & 0.711 $\pm$ 0.080 & mmag$^2$\\
	$\rho$ & Covariance decay timescale & $\No(0.036, 0.0036)$ & 0.0463 $\pm$ 0.0028 & days\\
	$t_0$ & Transit midpoint & $\No(2458411.8264, 0.1)$ & Table \ref{tbl:results} & BJD\\
	$\ln P$ & Orbital period & $\No(\ln 1.23913, 0.01)$ & Table \ref{tbl:results} & days\\
	$r/R$ & Planet-star radius ratio & $\Un(0.01, 0.4)$ & Table \ref{tbl:results} & -\\
	$b$ & Impact parameter & $\Un(0, 1+r/R)$ & Table \ref{tbl:results} & -\\
	\enddata
	\tablecomments{$\No(\mu, \sigma)$ denotes a normal distribution. $\Un(a, b)$ denotes a uniform distribution.}
\end{deluxetable}

We note that our reported model assumes a circular orbit (zero eccentricity). Due to the short orbital period of the planet, it is reasonable to expect that tidal dissipation has damped any initial amount of orbital eccentricity to an undetectably small level. Based on the work of \citet{Goldreich1966}, the expected tidal circularization timescale is close to 150,000 years (assuming the estimated bulk density from Section \ref{sec:results} and a specific dissipation function of $Q = 100$, which is appropriate for terrestrial planets), much shorter than the expected age of the system. Allowing the eccentricity $e$ and the angle of periastron passage $\omega$ to fluctuate produces a posterior probability distribution that is consistent with $e = 0$ at the cost of significantly broadening the posterior distributions of several other parameters (such as the impact parameter $b$) that alter the light curve in a similar way. Since an eccentricity much greater than zero would be inconsistent with the tidal circularization timescale, we keep $e$ fixed to 0 to constrain the other parameters better.

We proceed to tune and sample the model posterior distributions with \pymc. The MCMC sampling is done in parallel in 4 independent chains. Sampling from each chain begins with a burn-in phase of 1000 steps with an automatically tuned step size such that the acceptance fraction approaches 90\%, which facilitates convergence in complicated posterior distributions. We draw 1000 samples from each chain, for a total of 4000 samples. The sample distributions from each chain can be compared to each other to detect possible issues related to convergence. We detect no such problems: the resulting parameter distributions from the four chains are all consistent with one another. We report the means and the standard deviations of the modeled transit parameters from the 4000 samples in Table \ref{tbl:results} as well as the rest of the parameters (including hyperparameters) in Table \ref{tbl:params}. The results of the modeling are described in detail in Section \ref{sec:results}, and the modeled light curve can be seen overlaid on top of the \tess~and MEarth raw data in Figures \ref{fig:tessgp} and \ref{fig:mearth}, respectively.

\begin{deluxetable}{lcc}
    \tabletypesize{\footnotesize}
	\tablewidth{0pt}
	\tablecaption{System parameters for \starname\label{tbl:results}}
	\tablehead{
		Parameter & Values for \starname & Source\tablenotemark{a}
	}
	\startdata
	\multicolumn{3}{l}{\textbf{Stellar parameters}}\\
	Right ascension (J2000) & 05h 05min 14.4s & (1)\\
	Declination (J2000) & -47$^\circ$ 56' 15.5" & (1)\\
	\multirow{2}{*}{Proper motion (mas yr$^{-1}$)} & 
	    $\mu_\alpha = -66.09 \pm 0.08$ & \multirow{2}{*}{(1)}\\
	    & $\mu_\delta = 25.08 \pm 0.09$ & \\
	\multirow{6}{*}{Apparent brightness (mag)} &
	        $V_J = 14.492 \pm 0.03$ & (3)\\
	        & $R_{KC} = 13.115 \pm 0.03$ & (3)\\
	        & $I_{KC} = 11.402 \pm 0.03$ & (3)\\
	        & $J = 9.755 \pm 0.022$ & (2)\\
	        & $H = 9.170 \pm 0.022$ & (2)\\
	        & $K_S = 8.900 \pm 0.021$ & (2)\\
	Distance (pc) & 14.0078 $\pm$ 0.0088 & (1)\\
	Mass (\msun) & 0.159 $\pm$ 0.014 & (3)\\
	Radius (\rsun) & 0.1895 $\pm$ 0.0079 & (3)\\
	Luminosity (\lsun) & 0.00346 $\pm$ 0.00021 & (3)\\
	Fractional X-ray luminosity $L_X / L_{\rm bol}$ & 0.0028 & (3)\\
	X-ray flux\tablenotemark{b} (ergs s$^{-1}$ cm$^{-2}$) & $(1.60 \pm 0.56) \times 10^{-12}$ & (3)\\
	Effective temperature (K) & 3216 $\pm$ 83 & (3)\\
	Age (Gyr) & 0.1-2 Gyr & (3)\\
	Rotational period (days) & 0.72610 $\pm$ 0.00039 & (3)\\
	Projected rotation velocity (km/s) & 13.5 $\pm$ 1.5 & (3)\\
	\hline\hline
	Parameter & Values for \planetname & \\
	\hline
	\multicolumn{3}{l}{\textbf{Modeled transit parameters}}\\
	Orbital period $P$ (days) & 1.2391491 $\pm$ 0.0000017 & \\
	Eccentricity $e$ & 0 (fixed) & \\
	Time of mid-transit $t_{\rm T}$ (BJD) & 2458411.82601 $\pm$ 0.00046 & \\
	Impact parameter $b$ & 0.772 $\pm$ 0.029 & \\
	Planet-to-star radius ratio $r/R$ & 0.0436 $\pm$ 0.0012 & \\
	$a/R$ ratio & 13.90 $\pm$ 0.72 & \\
	\hline
	\multicolumn{3}{l}{\textbf{Derived planetary parameters}}\\
	Radius $r$ (\rearth) & 0.903 $\pm$ 0.052 & \\
	Semi-major axis $a$ (AU) & 0.01223 $\pm$ 0.00036 & \\
	Inclination $i$ (deg) & 86.80 $\pm$ 0.28 & \\
	\added{Bolometric} incident flux $S$ (\searth) & 23.4 $\pm$ 2.1 & \\
	Equilibrium temperature\tablenotemark{c} $T_{\rm eq}$ (K) & 611 $\pm$ 23 & \\
	\enddata
	\tablenotetext{a}{(1) \citet{Gaia2018}, (2) \citet{Skrutskie2006}, (3) This work.}
	\tablenotetext{b}{The X-ray flux is given in the \rosat~bandpass of 0.1-2.4 keV.}
	\tablenotetext{c}{The equilibrium temperature assumes a Bond albedo of 0. For an albedo of $A_{B}$, the reported temperature has to be multiplied by $(1-A_{B})^{1/4}$.}
\end{deluxetable}

\section{Discussion and conclusion}\label{sec:results}

\begin{figure}
    \includegraphics[width=\textwidth]{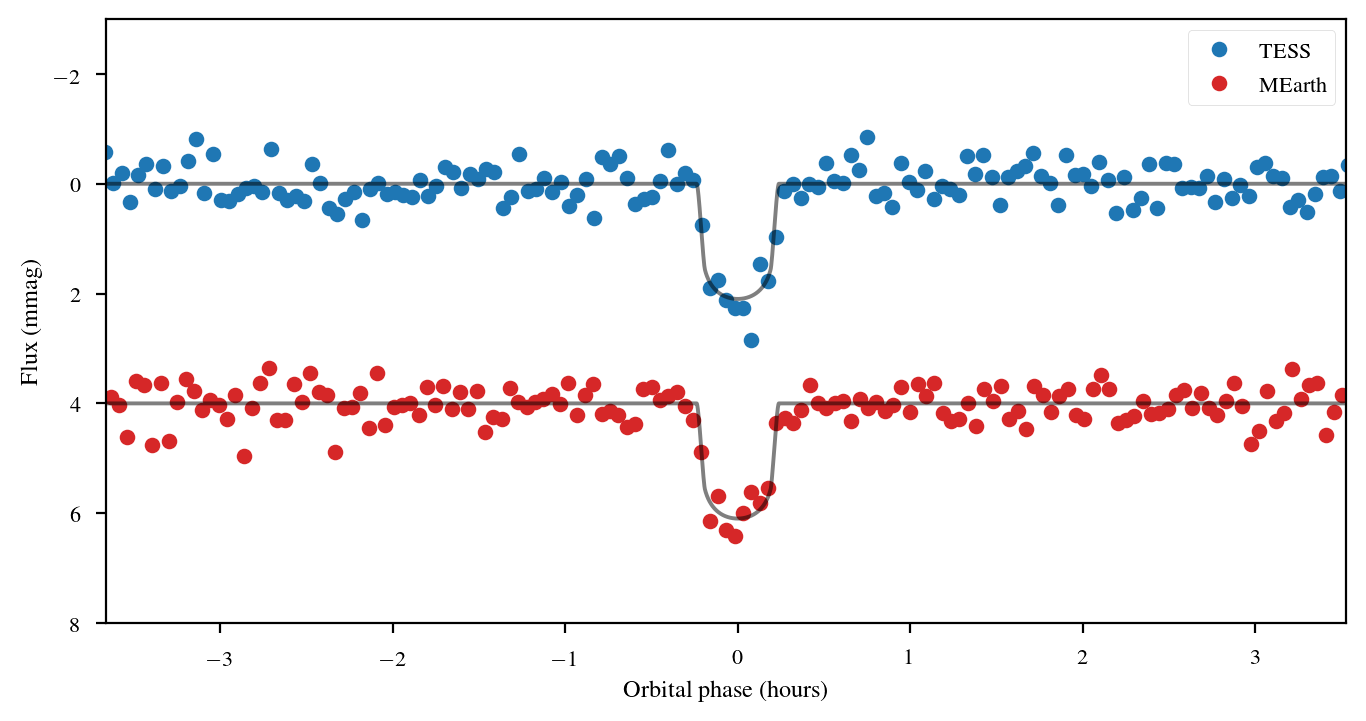}
    \caption{A phase-folded transit model of \planetname~from Section \ref{sec:modeling} overlaid on detrended \tess~and MEarth data. Both data sets have been binned to 2.88-minute intervals for visual clarity. The number of transits fitted is 50 for the TESS data and 9 for MEarth.}
    \label{fig:transits}
\end{figure}

\planetname~completes a trip around its rapidly rotating host star once every $P = 1.239149$ days ($\pm 170$ ms). The transits are not grazing with an impact parameter of $b = 0.772 \pm 0.029$, corresponding to an inclination angle of $i = 86.80\degree \pm 0.28\degree$. The planet has a radius of $r = 0.903 \pm 0.052$ \rearth, slightly less than that of Venus. Based on its small radius, the planet is likely to be terrestrial for the reasons outlined in Section \ref{sec:intro}. Using Earth's core mass fraction and a semi-empirical mass-radius relation for rocky planets by \citet{Zeng2016} yields an estimated mass of $m = 0.69 \pm 0.15$ \mearth~and a bulk density of $\rho = 5.2 \pm 1.4$ g cm$^{-3}$. This corresponds to a surface gravity of $g = 8.3 \pm 2.0$ m s$^{-2}$. Figure \ref{fig:transits} displays the predicted transits overlaid on detrended and co-added MEarth and \tess~data; the transit depth is consistent within 1$\sigma$ in either data set, and the residual noise levels are comparable.

\begin{figure}
    \includegraphics[width=\textwidth]{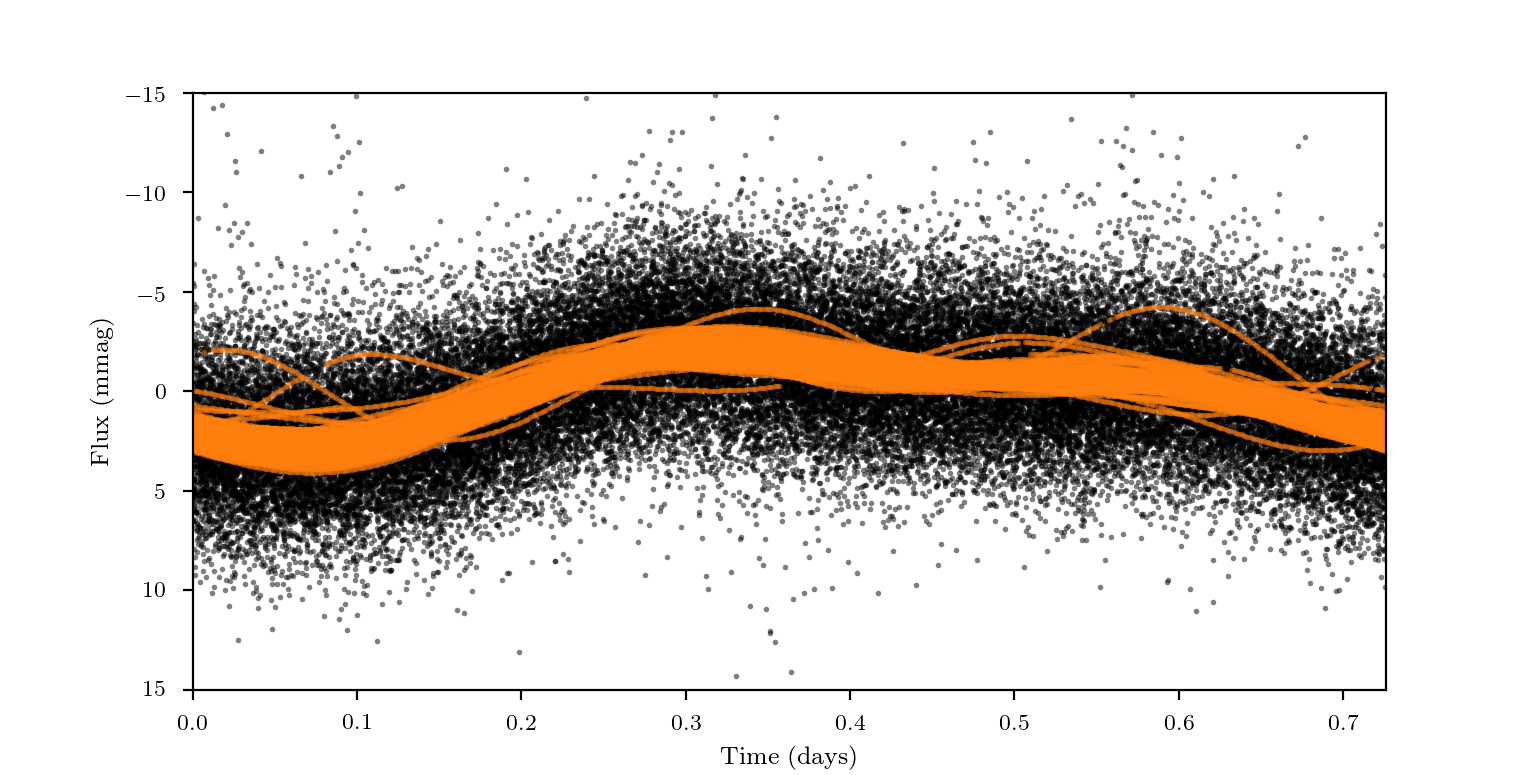}
    \caption{TESS photometry of \starname, phase-folded to the rotation period in Table \ref{tbl:results}. Solid orange lines overlay the data points, depicting the predicted rotational modulation for each individual observed stellar rotation. The orange curves overlap for the most part, implying that there is no evidence for long-term evolution in the modulation pattern, e.g. due to changes in the starspot coverage.}
    \label{fig:tessrot}
\end{figure}

The mass and radius of the star are not well constrained by the time series photometry alone and are almost entirely dictated by the prior distributions. The model finds a well-defined stellar rotation period of $\prot = 17.4264 \pm 0.0094$ hours. Even though the periodicity of the rotation can be established with great precision, the photometric modulation itself has multiple peaks per rotation, pointing to a heterogeneous distribution of spots across the photosphere. The shape of the modulation imprinted onto the \tess~light curve can be seen in Figure \ref{fig:tessrot}. However, there is no visual evidence of substantial evolution of the modulated signal over the two and a half months of data collection, facilitating the modeling necessary to isolate the transits of \planetname.

The combination of a slowly evolving starspot distribution and a short orbital period that allows for the observation of a large number of transits provides a rare opportunity to make use of \starname~to study the atmospheric composition and escape in small planets orbiting active M dwarfs. Out of the known planet systems within 15 pc of the Sun, similar conditions may perhaps be found only in the TRAPPIST-1 system \citep[TRAPPIST-1 has been estimated to spin with a period of 1.4 or 3.3 days;][]{Gillon2016,Luger2017} - however, \starname~is nearly 2 magnitudes brighter in the J band (and even more so towards the visible). In particular, rapidly rotating M dwarfs have more flares and CMEs, stronger stellar winds, and higher levels of X-ray and UV emission compared to slowly rotating M dwarfs, which likely lead to extensive atmospheric erosion \citep[and references therein]{Vida2017,Newton2018}. \citet{Wright2018} demonstrate a clear relationship between a star's Rossby number $\rm Ro = \prot/\tau$ (the ratio of the rotation period to the convective turnover time) and its coronal X-ray emission as a fraction of the bolometric luminosity. For stars with $\rm Ro > 0.14$, this relationship has a power-law slope, with smaller Rossby numbers corresponding to higher X-ray luminosities. Stars with $\rm Ro < 0.14$ (rapid rotators) have the highest levels of X-ray emission that remains saturated at a constant level of $L_{\rm X} / L_{\rm bol} \simeq 10^{-3}$. Using an empirically calibrated relation based on the $V_J-K_S$ color, we can calculate the convective turnover time of \starname~to be close to $\tau = 109$ days \citep[Eq. 5]{Wright2018}. This corresponds to a Rossby number of $\rm Ro = 0.007$, suggesting that \starname~is in the saturated high X-ray emission regime with an X-ray-to-bolometric luminosity ratio of $L_{\rm X} / L_{\rm bol} \simeq 10^{-3}$. This hypothesis is consistent with the X-ray detection of \starname~by \xmm~that yields $L_X / L_{\rm bol} = 0.0028$. Of the transiting planet hosts listed in Section \ref{sec:intro}, the only other star likely to be in the saturated regime is TRAPPIST-1 with $\rm Ro \approx 0.002-0.006$. This is based on an estimated convective turnover time $\tau = 582$ days that may be inaccurate as it was derived from stars with $V_J - K_S < 7.0$ \citep{Wright2018} whereas TRAPPIST-1 is substantially redder with $V_J - K_S = 8.5$ \citep{Gillon2017}. The bright X-ray flux of \starname~could present an opportunity to study transits in the X-ray to search for atmospheric loss, although we note that the atmospheric signature would likely be much smaller than the similar detection for HD 189733 b by \citet{Poppenhaeger2013}.

\begin{figure}
    \includegraphics[width=\textwidth]{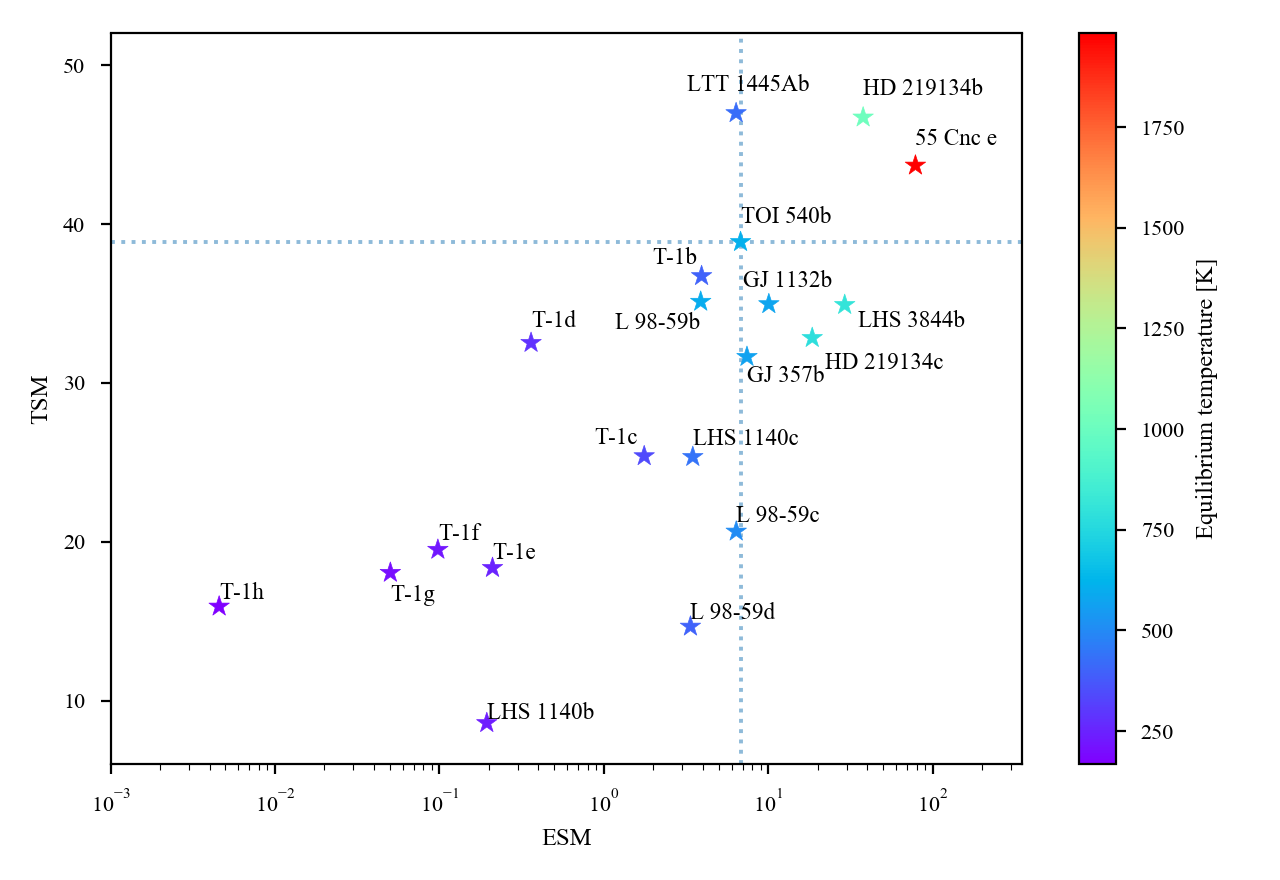}
    \caption{Estimated ESM and TSM values \citep{Kempton2018} for the known terrestrial transiting planets within 15 pc of the Sun, highlighting the most promising nearby targets for emission and transmission spectroscopy. Crucially, the two best targets in the top right (55 Cnc e and HD 219134 b) orbit larger K dwarfs and may not be detectable by \jwst~once a systematic noise floor is taken into account. The colorbar represents the equilibrium temperature of the planets assuming zero albedo and full day-night heat redistribution according to Eq. 3 of \citet{Kempton2018}. The TSM values were calculated with a scale factor of 0.190, calibrated for small planets. The dotted lines accentuate the location of \planetname. The values for all parameters were adopted from the publications listed in Section \ref{sec:intro}, and masses were estimated from \citet{Zeng2016} (with an Earth-like core mass fraction of 0.33) where not available. In addition, we adopted the updated mass measurements for TRAPPIST-1 (T-1) from \citet{Grimm2018}. Wherever missing from the publications listed in Section \ref{sec:intro}, $J$ and $K_S$-band fluxes were adopted from \citet{Ducati2002} and \citet{Skrutskie2006}.}
    \label{fig:tsm}
\end{figure}

The planet is likely to be hot: it receives 23.4 times the total radiation from its host star than Earth does from the Sun (and 3.5 times more than Mercury does), equating to a zero-albedo equilibrium temperature of $\teq = 611 \pm 23$ K. The high temperature, however, may make it more amenable to transmission and emission spectroscopy measurements. We calculate the transmission spectroscopy metric (TSM) and emission spectroscopy metric (ESM) from \citet{Kempton2018} for all of the nearby transiting terrestrial planets, listed in Section \ref{sec:intro}. The values are displayed in Figure \ref{fig:tsm}. Crucially, the two best targets in the top right of Figure \ref{fig:tsm} (55 Cnc e and HD 219134 b) orbit larger K dwarfs and may not be detectable by \jwst~once a systematic noise floor is taken into account. Using the scale factors from \citet{Kempton2018}, we obtain TSM = 38.9 and ESM = 6.8 for \planetname. The TSM value is above the threshold of 10 suggested by \citet{Kempton2018} and exceeds the TSM values of all but one of the known planets orbiting M dwarfs in Figure \ref{fig:tsm}. Therefore, \planetname~is a prime target for transmission spectroscopy to study a potential high mean molecular weight atmosphere, if indeed such an atmosphere can be retained in close proximity to an active M dwarf. Furthermore, the relatively high ESM value will likely qualify the planet for infrared photometry with \jwst~to detect or rule out the presence of atmosphere in as much as a single secondary eclipse \citep{Koll2019}. Finally, studies of the near-infrared or infrared phase curve could also tell us if \starname~has retained an atmosphere. A thermal phase curve study was used by \citet{Kreidberg2019} to rule out the presence of a thick atmosphere on LHS 3844 b, an ultra-short-period terrestrial planet orbiting a more evolved and less active M dwarf that has spun down to a rotation period of 128 days \citep{Vanderspek2019} but has a mass similar to that estimated for \starname. A similar thermal phase curve study for \starname~is promising.

We also cannot rule out the presence of additional transiting planets in this active system. We carried out a Box-Least Squares analysis \citep{Kovacs2002,Burke2006} as implemented in \citet{Ment2019} on the TESS light curve, but did not find sufficient evidence for additional transiting planets, consistent with the results of the SPOC’s search for additional planets. Considering the orbital inclination angle of \planetname, any co-planar transiting planets would be limited to orbital periods of 1.83 days or less, which would likely lead to dynamical instability given the orbital period of \planetname. Therefore, any additional planets around \starname~are likely to be non-transiting, or have an inclination substantially closer to 90\degree~than \planetname. Given the cumulative occurrence rate of $2.5 \pm 0.2$ small planets ($1-4$\rearth, $P<200$d) per M dwarf \citep{Dressing2015} that was derived from the \textit{Kepler} population, alternative methods such as transit timing variation (not detected for the 59 transits observed here) or RV studies (provided that enough observations can be accumulated to overcome the significant rotational broadening of spectral lines) may be fruitful to uncover more planets around \starname. In addition, high-resolution spectroscopy would allow for a precise modeling of line profiles during transit, yielding a direct measurement of stellar obliquity from Doppler tomography/the Rossiter-McLaughlin effect.

\acknowledgments
The MEarth team acknowledges funding from the David and Lucile Packard Fellowship for Science and Engineering (awarded to DC). This material is based on work supported by the National Science Foundation under grants AST-0807690, AST-1109468, AST-1004488 (Alan T. Waterman Award) and AST-1616624. This publication was made possible through the support of a grant from the John Templeton Foundation. The opinions expressed in this publication are those of the authors and do not necessarily reflect the views of the John Templeton Foundation. This material is based upon work supported by the National Aeronautics and Space Administration under Grant No. 80NSSC18K0476 issued through the XRP Program. AAM acknowledges support from the NSF Graduate Research Fellowship under Grant No. DGE1745303. RC is supported by a grant from the National Aeronautics and Space Administration in support of the TESS science mission. MRD is supported by CONICYT-PFCHA/Doctorado Nacional-21140646, Chile. JSJ is supported by funding from Fondecyt through grant 1201371 and partial support from CONICYT project Basal AFB-170002. BRA acknowledges the funding support from FONDECYT through grant 11181295. We acknowledge the use of public TESS Alert data from the pipelines at the TESS Science Office and at the TESS Science Processing Operations Center. Resources supporting this work were provided by the NASA High-End Computing (HEC) Program through the NASA Advanced Supercomputing (NAS) Division at Ames Research Center for the production of the SPOC data products. This research was made possible through the use of the AAVSO Photometric All-Sky Survey (APASS), funded by the Robert Martin Ayers Sciences Fund and NSF AST-1412587. This publication makes use of data products from the Two Micron All Sky Survey, which is a joint project of the University of Massachusetts and the Infrared Processing and Analysis Center/California Institute of Technology, funded by the National Aeronautics and Space Administration and the National Science Foundation. For securing the VRI photometry reported in this work, we thank RECONS (\url{www.recons.org}) members Andrew Couperus, Todd Henry, Wei-Chun Jao, and Eliot Vrijmoet. This work has made use of data from the European Space Agency (ESA) mission Gaia (\url{https://www.cosmos.esa.int/gaia}), processed by the Gaia Data Processing and Analysis Consortium (DPAC; \url{https://www.cosmos.esa.int/web/gaia/dpac/consortium}). Funding for the DPAC has been provided by national institutions, in particular the institutions participating in the Gaia Multilateral Agreement. This research has made use of data obtained from XMMSL2, the Second XMM-Newton Slew Survey Catalogue, produced by members of the XMM SOC, the EPIC consortium, and using work carried out in the context of the EXTraS project ("Exploring the X-ray Transient and variable Sky", funded from the EU's Seventh Framework Programme under grant agreement no. 607452). This research made use of \exoplanet~\citep{exoplanet} and its dependencies \citep{exoplanet:agol19, exoplanet:astropy13, exoplanet:astropy18,
exoplanet:foremanmackey17, exoplanet:foremanmackey18,
exoplanet:luger18, exoplanet:pymc3, exoplanet:theano}.

\facilities{TESS, MEarth, CTIO:1.5m (CHIRON), ESO:3.6m (HARPS), SOAR, XMM}

\bibliography{main}{}
\bibliographystyle{aasjournal}



\end{document}